# HS-Patch: A New Hermite Smart Bicubic Patch Modification

Vaclav Skala, Michal Smolik, Lukas Karlicek

*Abstract*—Bicubic four-sided patches are widely used in computer graphics, CAD/CAM systems etc. Their flexibility is high and enables to compress a surface description before final rendering. However, computer graphics hardware supports only triangular meshes. Therefore, four-sided bicubic patches are approximated by a triangular mesh. The border curves of a bicubic patch are of degree 3, while diagonal and anti-diagonal curves are of degree 6. Therefore the resulting shape and texturing depend on the actual mapping, i.e. how the tessellation of a bicubic patch is made.

The proposed new modification of the Hermite bicubic patch, the HS-patch, is a result of additional restriction put on the Hermite bicubic patch formulation – the diagonal and anti-diagonal curves are of degree 3. This requirement leads to a new Hermite based bicubic four-sided patch with 12 control points and another 4 control points, i.e. twist vectors, are computed from those 12 control points.

## I. Introduction

CUBIC parametric curves and surfaces are very often used for data interpolation or approximation. In the vast majority rectangular patches are used in engineering practice as they seem to be simple, easy to handle, compute and render (display). For rendering a rectangular patch is tessellated to a set of triangles. In some cases the definition domain is triangulated and users require smooth interpolation. In this case the mapping from triangles to parametric patches. However, this might lead to unexpected results as some edges of a triangular mesh will be mapped to curves of degree 3 and some of those to curves of degree 6.

In this paper we describe a new bicubic patch modification, called Hermite Smart (HS) patch. It is based on a Hermite bicubic patch on which some additional requirements are applied. This modification is motivated by engineering applications, in general. It is expected that the proposed HS-patch can be widely applied within GIS systems and geography applications as well.

## II. Problem Formulation

Parametric cubic curves and surfaces are described in many publications [1-7]. There are many different formulas for cubic curves and patches, e.g. Hermite, Bézier, Coons, B-spline etc., but generally diagonal curves of a bicubic rectangular patch are curves of degree 6. The proposed HS-Patch, derived from the Hermite form, has diagonal curves

The project was supported by the Ministry of Education of the Czech Republic, projects No.LH12181, No.LG13047 and SGS-2013-029.

Vaclav Skala, Michal Smolik and Lukas Karlicek are with Department of Computer Science and Engineering at Faculty of Applied Sciences, University of West Bohemia, Univerzitni 22, CZ 306 14 Plzen, Czech Republic. (web: http://www.VaclavSkala.eu)

of degree 3, i.e. curves for $v = u$ and $v = 1 - u$, while the original Hermite patch diagonal curves are of degree 6. Therefore the proposed HS-patch surface is "independent" of tessellation of the $u - v$ domain. It means that if any tessellation is used, all curves, i.e. boundary, diagonal and anti-diagonal curves are of degree 3. A cubic Hermite curve, Fig.1, can be described in a matrix form as

$$\boldsymbol{x}(t) = \boldsymbol{x}^T \boldsymbol{M}_H \, t \qquad \boldsymbol{M}_H = \begin{bmatrix} 2 & -3 & 0 & 1 \\ -2 & 3 & 0 & 0 \\ 1 & -2 & 1 & 0 \\ 1 & -1 & 0 & 0 \end{bmatrix} \qquad (1)$$

where: $\boldsymbol{x} = [x_1, x_2, x_3, x_4]^T$ is a vector of "control" values of a Hermite curve, $x_3 = \frac{dx_1}{dt}$ and $x_4 = \frac{dx_2}{dt}$, $\boldsymbol{t} = [t^3, t^2, t, 1]^T$, $t \in \langle 0, 1 \rangle$ is a parameter of the curve and $\boldsymbol{M}_H$ is the Hermite matrix.

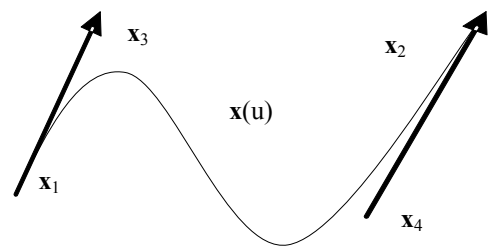

Fig.1 Hermite curve formulation

Generally we can write:

$$\boldsymbol{P}(t) = \boldsymbol{P}^T \boldsymbol{M}_H \, t \qquad (2)$$

where: $\boldsymbol{P}(t) = [x(t), y(t), z(t)]^T$

A bicubic Hermite patch, Fig.2, is described in a matrix form for the $x$-coordinate as

$$\boldsymbol{x}(u, v) = \boldsymbol{u}^T \boldsymbol{M}_H^T \, \boldsymbol{X} \, \boldsymbol{M}_H \, \boldsymbol{v} \qquad (3)$$

where: $\boldsymbol{X}$ is a matrix of "control" values of the Hermite cubic patch

$$\boldsymbol{X} = \begin{bmatrix} x_{11} & x_{12} & x_{13} & x_{14} \\ x_{21} & x_{22} & x_{23} & x_{24} \\ x_{31} & x_{32} & x_{33} & x_{34} \\ x_{41} & x_{42} & x_{43} & x_{44} \end{bmatrix} \quad \text{or} \quad \boldsymbol{X} = \begin{bmatrix} x_{ij} & \dfrac{\partial x_{ij}}{\partial v} \\ \dfrac{\partial x_{ij}}{\partial u} & \dfrac{\partial^2 x_{ij}}{\partial u \partial v} \end{bmatrix} \qquad (4)$$

where: i, j = 1,2,

$\boldsymbol{u}$, resp. $\boldsymbol{v}$ are vectors $\boldsymbol{u} = [u^3, u^2, u, 1]^T$, resp. $\boldsymbol{v} = [v^3, v^2, v, 1]^T$ and $u \in \langle 0, 1 \rangle$, resp. $v \in \langle 0, 1 \rangle$ are parameters of the patch.

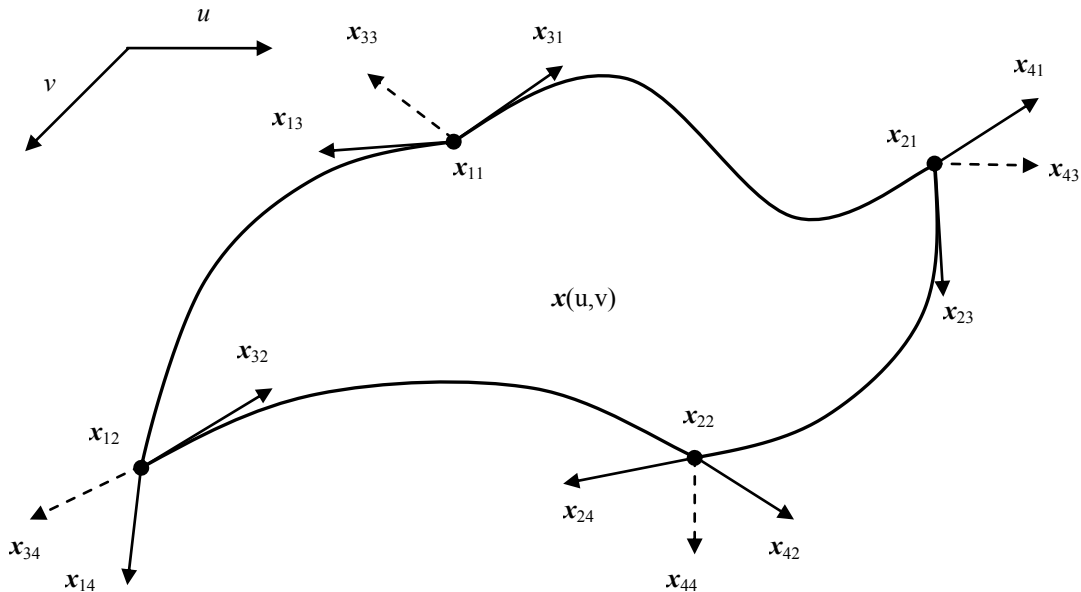

Fig.2 Hermite patch formulation

Similarly for $y$ and $z$ coordinates:

$$y(u,v) = \mathbf{u}^T \mathbf{M}_H^T \mathbf{Y} \mathbf{M}_H \mathbf{v} \qquad (5)$$
$$z(u,v) = \mathbf{u}^T \mathbf{M}_H^T \mathbf{Z} \mathbf{M}_H \mathbf{v}$$

It means that a rectangular Hermite patch is given by a matrix 4 x 4 of control values for each coordinate, i.e. by 3 x 16 = 48 values in $E^3$. From the definition of the Hermite patch it is clear, that boundary curves are cubic Hermite curves, i.e. curves of degree 3.

Generally we can rewrite the equations above as

$$\mathbf{P}(u,v) = \mathbf{u}^T \mathbf{M}_H^T \mathbf{P} \mathbf{M}_H \mathbf{v} \qquad (6)$$

where:

$$\mathbf{P}(u,v) = [x(u,v), y(u,v), z(u,v)]^T \qquad (7)$$

This equation is valid for other forms, like Bézier and B-Spline. They can be transformed linearly from one form to another and we can write for $x$ coordinate formally:

$$x(u,v) = \mathbf{u}^T \mathbf{M}_B^T \mathbf{X}_B \mathbf{M}_B \mathbf{v} \qquad (8)$$
$$x(u,v) = \mathbf{u}^T \mathbf{M}_S^T \mathbf{X}_S \mathbf{M}_S \mathbf{v}$$

where $\mathbf{X}_B$, resp. $\mathbf{X}_B$ are control points of the Bézier or B-spline forms.

Similarly for $y(u,v)$ and $z(u,v)$ coordinates. Of course, the $\mathbf{X}$ matrix has different interpretations. The $\mathbf{M}_B$ matrix for the Bézier, resp. $\mathbf{M}_S$ for the B-Spline forms are given as:

$$\mathbf{M}_B = \begin{bmatrix} -1 & 3 & -3 & 1 \\ 3 & -6 & 3 & 0 \\ -3 & 3 & 0 & 0 \\ 1 & 0 & 0 & 0 \end{bmatrix} \quad \mathbf{M}_S = \frac{1}{6}\begin{bmatrix} -1 & 3 & -3 & 1 \\ 3 & -6 & 0 & 4 \\ -3 & 3 & 3 & 1 \\ 1 & 0 & 0 & 0 \end{bmatrix} \qquad (9)$$

However, there are many applications, where a rectangular or a triangular mesh is used in the $u - v$ domain. Sometimes $x$, resp. $y$ values are taken as $u$, resp. $v$ parameters and only $z$ value is interpolated/approximated as $z = f(x,y)$.

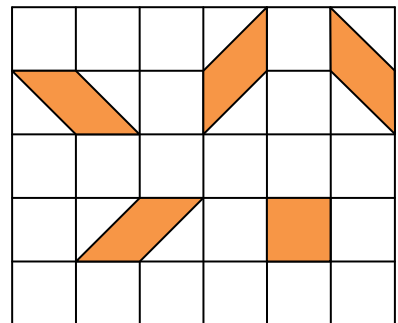

Fig.3 Different tessellation of u-v domain for the given corner points

There are many practical reasons why patches, i.e. the $u - v$ domain, are tessellated to a triangular mesh, let us present just some of them:

- A plane in $E^3$ is defined by three points, so the 4[th] point is not generally on the plane (due to computer limited precision it is nearly always invalid even if the point theoretically lies on the plane).
- The given $u - v$ rectangular domain mesh can be tessellated in different ways to a triangular mesh, in general, using different patterns (see Fig.3).

If a rectangular Hermite cubic patch is used, then the diagonal curves, i.e. $v = u$ and $v = 1 - u$, are of degree 6.

Especially last two points are very important as the final surface depends on tessellation and some curves might be of degree 6. This is not acceptable for some applications. There is a natural question:

"Why some curves are of degree 3 and some are of degree 6 (when fixing $u = f(v)$)?".

If this feature is not controlled carefully it could lead to critical, sometimes even fatal, situations.

Understanding this, we exposed a specific restriction to the Hermite patch as curves for $v = u$ and $v = 1 - u$ must be of degree 3 as the patch boundary curves. This requirement has resulted into new modification of the Hermite cubic patch, called Hermite Smart Patch (HS-patch), described below.

### III. PROPOSED HS-PATCH

Let us consider the Hermite patch on which we put some restrictions given by the requirement that diagonal curves, i.e. for $v = u$ and $= 1 - u$, are to be of degree 3.
The Hermite patch is given in the matrix form as

$$x(u,v) = \boldsymbol{u}^T \boldsymbol{M}_H^T \boldsymbol{X} \boldsymbol{M}_H \boldsymbol{v} \qquad (10)$$

The restrictions for the proposed HS-patch are:
$x(u,v)$ for $v = u$ is a curve of degree 3, it means that $x(u) = \boldsymbol{u}^T \boldsymbol{M}_H^T \boldsymbol{X} \boldsymbol{M}_H \boldsymbol{u}$ is a curve of degree 3.
We can write

$$x(u) = \boldsymbol{u}^T \boldsymbol{R}_1 \boldsymbol{u} \qquad (11)$$

where: $\boldsymbol{R}_1 = \boldsymbol{M}_H^T \boldsymbol{X} \boldsymbol{M}_H$ and $x(u,v)$ for $v = 1 - u$ is a curve of degree 3. It means that $x(u) = \boldsymbol{u}^T \boldsymbol{M}_H^T \boldsymbol{X} \boldsymbol{M}_H \boldsymbol{T} \boldsymbol{u}$ is a curve of degree 3, where:

$$\boldsymbol{X}_B = [(1-u)^3 \quad (1-u)^2 \quad 1-u \quad 1]^T$$

$$= \begin{bmatrix} -u^3 + 3u^2 - 3u + 1 \\ u^2 - 2u + 1 \\ -u + 1 \\ 1 \end{bmatrix}$$

$$= \begin{bmatrix} -1 & 3 & -3 & 1 \\ 0 & 1 & -2 & 1 \\ 0 & 0 & -1 & 1 \\ 0 & 0 & 0 & 1 \end{bmatrix} \begin{bmatrix} u^3 \\ u^2 \\ u \\ 1 \end{bmatrix} = \boldsymbol{T} \boldsymbol{u} \qquad (12)$$

We can write:

$$x(u) = \boldsymbol{u}^T \boldsymbol{R}_2 \boldsymbol{u} \qquad (13)$$

where:

$$\boldsymbol{R}_2 = \boldsymbol{M}_H^T \boldsymbol{X} \boldsymbol{M}_H \boldsymbol{T} \qquad (14)$$

The Hermite diagonal curve is in both cases defined as

For the case 1, i.e. $u = v$ we get

$$x(u) = \sum_{i,j=1}^{4} r_{ij} u^{4-i} u^{4-j} = r_{11} u^6 + (r_{12} + r_{21}) u^5 \\
+ (r_{13} + r_{22} + r_{31}) u^4 \\
+ (r_{14} + r_{23} + r_{32} + r_{41}) u^3 \\
+ (r_{24} + r_{33} + r_{42}) u^2 \\
+ (r_{34} + r_{43}) u + r_{44} \qquad (15)$$

$$x(u) = \sum_{k=0}^{6} a_k u^k = \boldsymbol{a}^T \boldsymbol{u}$$

where:

$$\boldsymbol{u} = [u^6, \ u^5, \ u^4, \ u^3, \ u^2, \ u, \ 1]^T \\
\boldsymbol{a} = [\, r_{11}, \ r_{12} + r_{21}, \ r_{13} + r_{22} + r_{31}, \ r_{14} + r_{23} + r_{32} \\
+ r_{41}, r_{24} + r_{33} + r_{42}, r_{34} \\
+ r_{43}, r_{44}\,]^T \qquad (16)$$

We have 16 equations giving the relations between $x_{ij}$ and $r_{ij}$ for both cases, i.e. $\boldsymbol{R}_1$ and $\boldsymbol{R}_2$, that form a matrix relation, which expresses how the control values $x_{ij}$ form the coefficients in the $r_{ij}$

$$\boldsymbol{\rho} = \boldsymbol{\Omega}\, \boldsymbol{\xi} \qquad (17)$$

where: $\boldsymbol{\rho}$ is a vector of coefficients of the matrix $\boldsymbol{R}$ and $\boldsymbol{\xi}$ is a vector of control points in the matrix $\boldsymbol{X}$.
We can write

$$a_i = \sum_{j=1}^{16} \lambda_{ij} \xi_j = \boldsymbol{\lambda}_i^T \boldsymbol{\xi} \qquad (18)$$

for $i = 1, \dots, 6$.
As we require the diagonal curves to be of degree 3, we can write conditions for that as:

$$r_{11} = 0;\ r_{12} + r_{21} = 0;\ r_{13} + r_{22} + r_{31} = 0 \qquad (19)$$

For the case 1, the matrix $\boldsymbol{R}_1$ is used, for the case 2, the matrix $\boldsymbol{R}_2$ is used.

$$\boldsymbol{\xi} = [\, x_{11},\ x_{12},\ x_{13},\ x_{14},\ x_{21},\ x_{22},\ x_{23},\ x_{24},\ x_{31},\ x_{32},\ x_{33},\ x_{34},\ x_{41},\ x_{42},\ x_{43},\ x_{44}\,]^T$$

$$\boldsymbol{\Omega}_1 = \begin{bmatrix}
4 & -4 & 2 & 2 & -4 & 4 & -2 & -2 & 2 & -2 & 1 & 1 & 2 & -2 & 1 & 1 \\
-6 & 6 & -4 & -2 & 6 & -6 & 4 & 2 & -3 & 3 & -2 & -1 & -3 & 3 & -2 & -1 \\
. & . & 2 & . & . & . & -2 & . & . & . & 1 & . & . & . & 1 & . \\
2 & . & . & . & -2 & . & . & . & 1 & . & . & . & 1 & . & . & . \\
-6 & 6 & -3 & -3 & 6 & -6 & 3 & 3 & -4 & 4 & -2 & -2 & -2 & 2 & -1 & -1 \\
9 & -9 & 6 & 3 & -9 & 9 & -6 & -3 & 6 & -6 & 4 & 2 & 3 & -3 & 2 & 1 \\
. & . & -3 & . & . & . & 3 & . & . & . & -2 & . & . & . & -1 & . \\
-3 & . & . & . & 3 & . & . & . & -2 & . & . & . & -1 & . & . & . \\
. & . & . & . & . & . & . & . & 2 & -2 & 1 & 1 & . & . & . & . \\
. & . & . & . & . & . & . & . & -3 & 3 & -2 & -1 & . & . & . & . \\
. & . & . & . & . & . & . & . & . & . & 1 & . & . & . & . & . \\
. & . & . & . & . & . & . & . & 1 & . & . & . & . & . & . & . \\
2 & -2 & 1 & 1 & . & . & . & . & . & . & . & . & . & . & . & . \\
-3 & 3 & -2 & -1 & . & . & . & . & . & . & . & . & . & . & . & . \\
. & . & 1 & . & . & . & . & . & . & . & . & . & . & . & . & . \\
1 & . & . & . & . & . & . & . & . & . & . & . & . & . & . & .
\end{bmatrix} \begin{bmatrix} r_{11} \\ r_{12} \\ r_{13} \\ r_{14} \\ r_{21} \\ r_{22} \\ r_{23} \\ r_{24} \\ r_{31} \\ r_{32} \\ r_{33} \\ r_{34} \\ r_{41} \\ r_{42} \\ r_{43} \\ r_{44} \end{bmatrix} \qquad (20)$$

For the case 2, i.e. $v = 1 - u$ we get

$$\xi = [\, x_{11},\ x_{12},\ x_{13},\ x_{14},\ x_{21},\ x_{22},\ x_{23},\ x_{24},\ x_{31},\ x_{32},\ x_{33},\ x_{34},\ x_{41},\ x_{42},\ x_{43},\ x_{44}\,]^T$$

$$\Omega_2 = \begin{bmatrix}
4 & -4 & 2 & 2 & -4 & 4 & -2 & -2 & 2 & -2 & 1 & 1 & 2 & -2 & 1 & 1 \\
-6 & 6 & -4 & -2 & 6 & -6 & 4 & 2 & -3 & 3 & -2 & -1 & -3 & 3 & -2 & -1 \\
. & . & . & -2 & . & . & . & -2 & . & . & . & -1 & . & . & . & -1 \\
. & 2 & . & . & . & -2 & . & . & . & 1 & . & . & . & 1 & . & . \\
6 & -6 & 3 & 3 & -6 & 6 & -3 & -3 & 4 & -4 & 2 & 2 & 2 & -2 & 1 & 1 \\
-9 & 9 & -3 & -6 & 9 & -9 & 3 & 6 & -6 & 6 & -2 & -4 & -3 & 3 & -1 & -2 \\
. & . & -3 & . & . & . & 3 & . & . & . & -2 & . & . & . & -1 & . \\
. & -3 & . & . & . & 3 & . & . & . & -2 & . & . & . & -1 & . & . \\
. & . & . & . & . & . & . & . & -2 & 2 & -1 & -1 & . & . & . & . \\
. & . & . & . & . & . & . & . & 3 & -3 & 1 & 2 & . & . & . & . \\
. & . & . & . & . & . & . & . & . & . & -1 & . & . & . & . & . \\
. & . & . & . & . & . & . & . & 1 & . & . & . & . & . & . & . \\
-2 & 2 & -1 & -1 & . & . & . & . & . & . & . & . & . & . & . & . \\
3 & -3 & 1 & 2 & . & . & . & . & . & . & . & . & . & . & . & . \\
. & . & . & -1 & . & . & . & . & . & . & . & . & . & . & . & . \\
. & 1 & . & . & . & . & . & . & . & . & . & . & . & . & . & .
\end{bmatrix}
\begin{bmatrix} r_{11} \\ r_{12} \\ r_{13} \\ r_{14} \\ r_{21} \\ r_{22} \\ r_{23} \\ r_{24} \\ r_{31} \\ r_{32} \\ r_{33} \\ r_{34} \\ r_{41} \\ r_{42} \\ r_{43} \\ r_{44} \end{bmatrix} \quad (21)$$

$$\xi = [\, x_{11},\ x_{12},\ x_{21},\ x_{22},\ x_{13},\ x_{14},\ x_{23},\ x_{24},\ x_{31},\ x_{32},\ x_{33},\ x_{34},\ x_{41},\ x_{42},\ x_{43},\ x_{44}\,]^T$$

$$\Lambda = \begin{bmatrix}
4 & -4 & 2 & 2 & -4 & 4 & -2 & -2 & 2 & -2 & 1 & 1 & 2 & -2 & 1 & 1 \\
-12 & 12 & -7 & -5 & 12 & -12 & 7 & 5 & -7 & 7 & -4 & -3 & -5 & 5 & -3 & -2 \\
9 & -9 & 8 & 3 & -9 & 9 & -8 & -3 & 8 & -8 & 6 & 3 & 3 & -3 & 3 & 1 \\
-4 & 4 & -2 & -2 & 4 & -4 & 2 & 2 & -2 & 2 & -1 & -1 & -2 & 2 & -1 & -1 \\
12 & -12 & 5 & 7 & -12 & 12 & -5 & -7 & 7 & -7 & 3 & 4 & 5 & -5 & 2 & 3 \\
-9 & 9 & -3 & -8 & 9 & -9 & 3 & 8 & -8 & 8 & -3 & -6 & -3 & 3 & -1 & -3
\end{bmatrix} \quad (22)$$

From those conditions we get a system of linear equations:

$$\Lambda\,\xi = 0 \quad (23)$$

where the first three rows of the matrix $\Lambda$ (22) are taken for the case 1, i.e. related to the matrix $R_1$, and last three rows are taken for the case 2, i.e. related to the matrix $R_2$.

Now we can write the equivalent to the equation $\Lambda\,\xi = 0$ as:

$$[\Lambda_2 \quad \Lambda_1]\begin{bmatrix} \xi_2 \\ \xi_1 \end{bmatrix} = 0 \quad (24)$$

As $\xi_1$ are given values (corner points of the patch) we can write:

$$\Lambda_2 \xi_2 = -\Lambda_1 \xi_1 \quad (25)$$

The rank of the matrix rank$(\Lambda) = 5$, which means that we have to respect some restrictions generally imposed on the control points of the HS-patch. As the corner points are given by a user, the tangent and twist vectors are tied together with a relation. It can be seen that the vector $\xi$ is actually composed from values that are fixed (corner points are usually given) and by values, that can be considered as "free", but have to fulfil some additional condition(s). Let us explore this condition more in detail, now.

The equation $\Lambda\,\xi = 0$ can be rewritten as corner points are given.

Let us define vectors $\xi_1$ and $\xi_2$, i.e. the corner points of the patch $\xi_1$ and tangent and twist vectors of the patch $\xi_2$, and matrices $\Lambda_1$ and $\Lambda_2$ as

$$\xi_1 = [\, x_{11},\ x_{12},\ x_{21},\ x_{22}\,]^T \qquad \xi_2 = [\, x_{13},\ x_{14},\ x_{23},\ x_{24},\ x_{31},\ x_{32},\ x_{33},\ x_{34},\ x_{41},\ x_{42},\ x_{43},\ x_{44}\,]^T$$

$$\Lambda_1 = \begin{bmatrix} 4 & -4 & -4 & 4 \\ -12 & 12 & 12 & -12 \\ 9 & -9 & -9 & 9 \\ -4 & 4 & 4 & -4 \\ 12 & -12 & -12 & 12 \\ -9 & 9 & 9 & -9 \end{bmatrix} \qquad \Lambda_2 = \begin{bmatrix} 2 & 2 & -2 & -2 & 2 & -2 & 1 & 1 & 2 & -2 & 1 & 1 \\ -7 & -5 & 7 & 5 & -7 & 7 & -4 & -3 & -5 & 5 & -3 & -2 \\ 8 & 3 & -8 & -3 & 8 & -8 & 6 & 3 & 3 & -3 & 3 & 1 \\ -2 & -2 & 2 & 2 & -2 & 2 & -1 & -1 & -2 & 2 & -1 & -1 \\ 5 & 7 & -5 & -7 & 7 & -7 & 3 & 4 & 5 & -5 & 2 & 3 \\ -3 & -8 & 3 & 8 & -8 & 8 & -3 & -6 & -3 & 3 & -1 & -3 \end{bmatrix} \quad (26)$$

As we can write the equivalent to the equation $\Lambda\,\xi = 0$ as:

$$[\Lambda_2 \quad \Lambda_1]\begin{bmatrix} \xi_2 \\ \xi_1 \end{bmatrix} = 0 \quad (27)$$

As $\xi_1$ are given values (corner points of the patch) we can write:

$$\Lambda_2 \xi_2 = -\Lambda_1 \xi_1 \tag{28}$$

Rewriting and reducing the system of equations above, we get:

$$\begin{bmatrix} x_{13} & x_{14} & x_{23} & x_{24} & x_{31} & x_{32} & x_{33} & x_{34} & x_{41} & x_{42} & x_{43} & x_{44} \\ 1 & . & -1 & . & . & . & . & . & 1 & -1 & 1 & . \\ . & 1 & . & -1 & . & . & . & . & 1 & -1 & . & 1 \\ . & . & . & . & 1 & -1 & . & . & -1 & 1 & -1 & -1 \\ . & . & . & . & . & . & 1 & . & . & . & . & 1 \\ . & . & . & . & . & . & . & 1 & . & . & 1 & . \end{bmatrix} \xi_2 = \begin{bmatrix} x_{11} & x_{12} & x_{21} & x_{22} \\ -1 & 1 & 1 & -1 \\ -1 & 1 & 1 & -1 \\ -2 & 2 & 2 & -2 \\ 2 & -2 & -2 & 2 \\ 2 & -2 & -2 & 2 \end{bmatrix} \xi_1 \tag{29}$$

From this equation (last two rows) we can see that the twist values of the patch must fulfil the following conditions:

$$\begin{aligned} x_{33} + x_{44} &= 2\varphi \\ x_{34} + x_{43} &= 2\varphi \\ \varphi &= x_{11} - x_{12} - x_{21} + x_{22} \end{aligned} \tag{30}$$

where $\varphi$ is given by the corner points of the bicubic patch.

We can define two parameters $\alpha$ and $\beta$ (actually barycentric coordinates) as follows:

$$2\varphi\,\alpha = x_{44} \qquad 2\varphi\,(1-\alpha) = x_{33} \tag{31}$$

and

$$2\varphi\,\beta = x_{43} \qquad 2\varphi\,(1-\beta) = x_{34} \tag{32}$$

It means that the twist vectors are determined by $\alpha$ and $\beta$ values and by the corner points.

Now we have the following equations to be solved

1$^{st}$ row $\quad b + x_{43} = b + 2\beta\varphi = -\varphi$,  
$\qquad\qquad b = x_{13} - x_{23} + x_{41} - x_{42}$  
2$^{nd}$ row $\quad a + x_{44} = a + 2\alpha\varphi = -\varphi$,  
$\qquad\qquad a = x_{14} - x_{24} + x_{41} - x_{42}$  
3$^{rd}$ row  
$\qquad c - x_{43} - x_{44} = c - 2(\alpha+\beta)\varphi = -2\varphi$  
$\qquad c = x_{31} - x_{32} - x_{41} + x_{42}$ (33)

Then

$$\beta = -\frac{b+\varphi}{2\varphi} \qquad \alpha = -\frac{a+\varphi}{2\varphi} \tag{34}$$

$$\begin{aligned} -2\varphi &= c - 2\varphi\alpha - 2\varphi\beta \\ &= c + 2\varphi\frac{a+\varphi}{2\varphi} + 2\varphi\frac{b+\varphi}{2\varphi} \\ -2\varphi &= c + a + \varphi + b + \varphi \\ a + b + c &= -4\varphi \end{aligned} \tag{35}$$

Expressing $\alpha$ and $\beta$ from the first two equations we get an equation (constraint) for the control values of the Hermite form that is actually the HS-patch as:

$$x_{31} - x_{32} + x_{41} - x_{42} + x_{14} - x_{24} - x_{23} + x_{13} = -4\varphi$$

i.e. (36)

$$x_{31} - x_{32} + x_{41} - x_{42} + x_{14} - x_{24} - x_{23} + x_{13} = -4[x_{11} - x_{21} - x_{12} + x_{22}]$$

This result should be read as follows:
- The HS-patch control points are determined parametrically. The control points (tangent vectors) of the border curves have to fulfil the condition above.
- Twist vectors of the HS-patch are controlled by values $\alpha$ and $\beta$ that are determined actually by the control points that form the patch boundary.

IV. REPRESENTATION CONVERSION

A parametric curve can be represented in the Hermite form as

$$\boldsymbol{P}(t) = \boldsymbol{P}_H^T \boldsymbol{M}_H\, t \tag{37}$$

where: $\boldsymbol{P}_H^T$ are control points in the Hermite form or in the Bézier form as

$$\boldsymbol{P}(t) = \boldsymbol{P}_B^T \boldsymbol{M}_B\, t \tag{38}$$

where: $\boldsymbol{P}_B^T$ are control points in the Bézier form. If in the both cases the curve is the same we can write

$$\boldsymbol{P}_B^T \boldsymbol{M}_B\, t = \boldsymbol{P}_H^T \boldsymbol{M}_H\, t \qquad \boldsymbol{P}_B^T \boldsymbol{M}_B = \boldsymbol{P}_H^T \boldsymbol{M}_H \tag{39}$$

and therefore

$$\boldsymbol{P}_H^T = \boldsymbol{P}_B^T \boldsymbol{M}_B \boldsymbol{M}_H^{-1} = \boldsymbol{P}_B^T \boldsymbol{M}_{BH} \tag{40}$$

where: $\boldsymbol{M}_{BH}$ is a transformation matrix from the Bézier form to the Hermite form; similarly for the B-Spline and other forms. We can write for Bézier form:

$$\boldsymbol{M}_{BH} = \begin{bmatrix} 1 & 0 & -3 & 0 \\ 0 & 0 & 3 & 0 \\ 0 & 0 & 0 & 3 \\ 0 & 1 & 0 & 3 \end{bmatrix} \qquad \boldsymbol{M}_{HB} = \frac{1}{3}\begin{bmatrix} 3 & 3 & 0 & 0 \\ 0 & 0 & 3 & 3 \\ 0 & 1 & 0 & 0 \\ 0 & 0 & -1 & 0 \end{bmatrix} \tag{41}$$

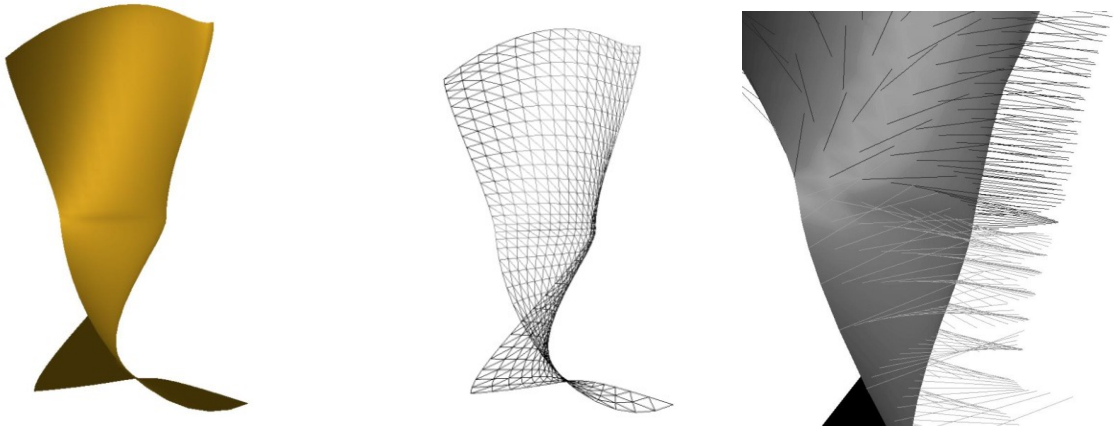

Fig.4 Two joined patches rendered, mesh and normal vectors

and for the B-Spline:

$$M_{SH} = \frac{1}{3}\begin{bmatrix} 1 & 0 & -3 & 0 \\ 4 & 1 & 0 & -3 \\ 1 & 4 & 3 & 0 \\ 0 & 1 & 0 & 3 \end{bmatrix} \quad M_{HS} = \frac{1}{3}\begin{bmatrix} -3 & 6 & -3 & 6 \\ 6 & -3 & 6 & -3 \\ -7 & 2 & -1 & 2 \\ -2 & 1 & -2 & 7 \end{bmatrix} \quad (42)$$

Derived conditions for the Bézier and B-Spline bicubic patches are not as simple as in the case of the Hermite form. Therefore the Hermite form is used as a fundamental form to which the data are transferred and in which the patches connection is made. Then the results are translated back to the original form if needed. The same approach was used for the Utah Teapot, see.Fig.7-10.

## V. LIMITATIONS

Conditions set for the diagonal and anti-diagonal curves causes some limitations that are not severe in the practical cases, but might cause some implementation problems. In this case the 4-sided patch must be split to five 4-sided patches by inserting 4 points. This special case is actually caused by the recently presented condition

$$x_{31} - x_{32} + x_{41} - x_{42} + x_{14} - x_{24} - x_{23} + x_{13} = -4[x_{11} - x_{21} - x_{12} + x_{22}] \quad (43)$$

Such situation can be demonstrated by a simple example. Let us consider a rectangular 4-sided patch in the $x$-$y$ plane and with $z = 0$ having corners A, B, C and D. All derivatives are $\partial z/\partial u = 0$ and $\partial z/\partial v = 0$ in those corners. Now if the corner D is moved to a position $z = 1$ than this situation cannot be represented by a single HS-patch. However if new 4 points are inserted we get actually five 4-sided patches instead of the original one. The new inserted points actually form a new patch inside if the original and the corner points of it are connected with the original corner points.

## VI. EXPERIMENTAL RESULTS

The experiments carried out proved that the proposed HS-Patch has reasonable geometric and user friendly properties.

To join HS-Patches together a similar approach can be taken as for the standard Hermite rectangular patch. There is just a small complication as the equation:

$$x_{31} - x_{32} + x_{41} - x_{42} + x_{14} - x_{24} - x_{23} + x_{13} = -4[x_{11} - x_{21} - x_{12} + x_{22}] \quad (44)$$

has to be respected and kept valid. It has just influence to a curvature of a surface of the neighbours.

The experiments made proved that it is possible to join HS-Patches smoothly. To prove additional basic properties of the proposed HS-Patch we used ½ of a cube.

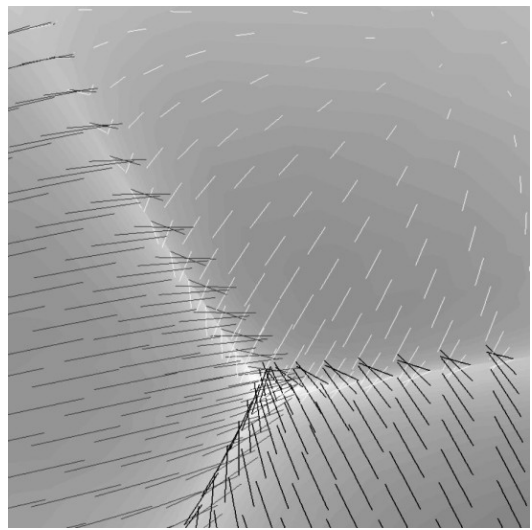

Fig.5 Joined patches of ½ cube – normal vectors

However, there was a severe problem detected, when a vertex of the mesh is shared by three patches. A solution can be found in [15]. In some cases it was difficult to keep $C^1$ continuity, Fig.5.

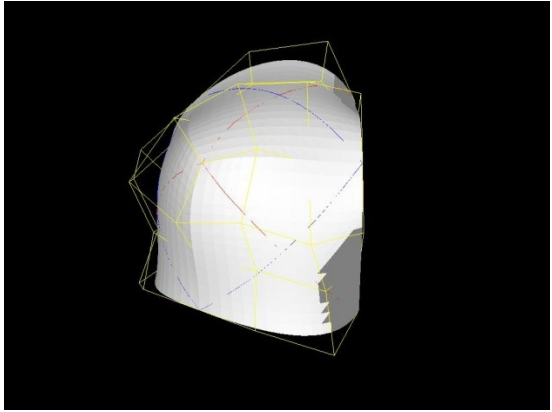

Fig. 6 Joined patches of ½ cube – rendered surface

Fig.8-10 presents the Utah Teapot modeled by the HS-patches and rendered. The edges of patches were highlighted to present the patches borders, the patches connections are $C^1$ or $G^1$ continuous similarly to standard patches connections. However it should be noted that the top of the cover there is only $C^0$ continuity and presents current semi limitation of the HS-patch.

It can be easily overcome by splitting the HS-patch horizontally into three patches, see Fig.7.

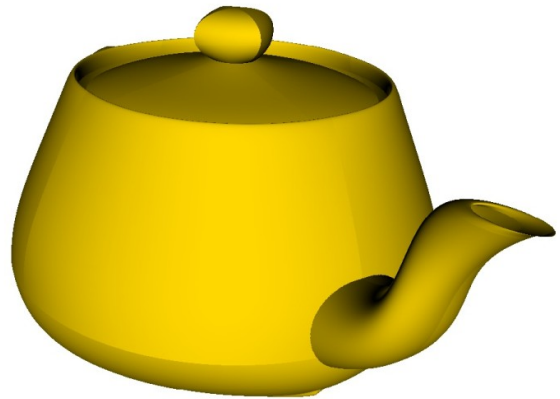

Fig.8 The Utah Teapot rendered by HS-Patch $C^1$ & $G^1$

Borders of patches are made explicitly visible as they have been rendered independently as a triangular mesh patch by patch. The surface is actually smooth.

At Fig.8 can be seen also violation of the condition for tangent vectors, while at Fig.9 and Fig.10 the top of the pot is split to more patches.

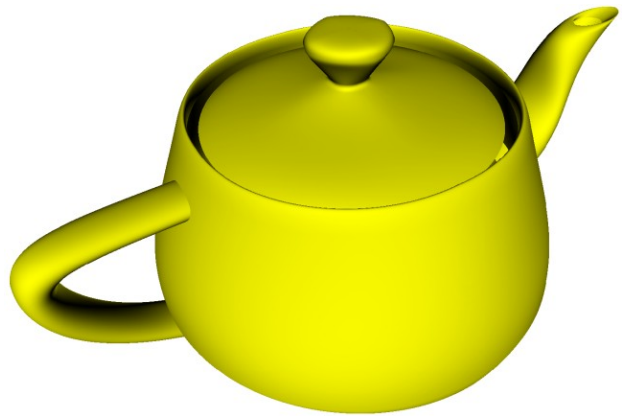

Fig.9 The Utah Teapot rendered by HS-Patch $C^1$ & $G^1$

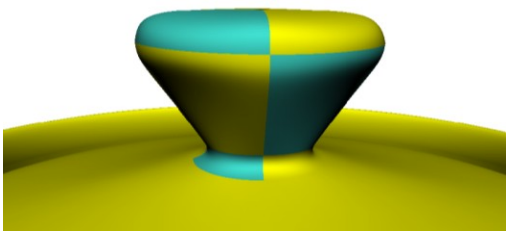

Fig.7 Splitted top of the cover of the Utah Teapot rendered by HS-Patch $C^1$ & $G^1$

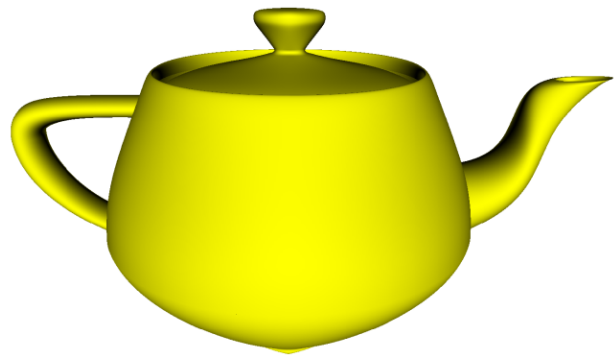

Fig.10 The Utah Teapot rendered by HS-Patch $C^1$ & $G^1$

A reader can find details on recent formulation of the S-Patch based approach as follows: the Hermite form [14] and S-Patch modifications for the Bézier, B-Spline and Catmul-Rom formulations [15]. The early formulation [14], [17] of the BS-patch was modified for the quadratic case by Kolcun [9].

## VII. CONCLUSION

We have described and derived a new modification of the Hermite bicubic patch. The main advantages of the proposed HS-patch are:
- Twist vectors are determined by the equation 1 due to the restriction put on diagonal curves as we require degree 3.
- Both diagonal curves are cubic curves, i.e. curves of degree 3.
- Different tessellations of $u - v$ domain and conversion to triangles do not change the degree of border and diagonal curves.
- Curves (boundary and diagonal) are of degree 3, less operations are needed as the computed polynomial is of degree 3.
- The given $u - v$ domain can be tessellated in different ways to four sided mesh and to triangular meshes for rendering using different tessellations.

It should be noted that one additional condition to be kept valid.

The future work will be devoted to:
- Exploration how the HS-patch formulation should be modified for cases when a vertex of the mesh is shared by 3 or 5 patches. New general conditions have been already derived [15] based on equivalence of solution of linear equations and generalized cross (outer) product.
- Analysis of the HS-patch properties (smoothness, curvature etc.)
- Methods for users interface implementation including user interaction and manipulation in surface design.
- Derivation of HS-patches conditions for other patch types.

ACKNOWLEDGMENT

The authors would like to express thanks to colleagues at the University of West Bohemia, Plzen for challenging this work, for many comments and hints they have made. Special thanks belong to Vit Ondracka from the University of West Bohemia for verification of some equations, to Alexej Kolcun from the Institute of Geonics, Academy of Sciences of the Czech Republic and to Rongjiang Pan, Shandong University, China and to Marc Daniel, LSIS, France for their comments and suggestions. Also comments by anonymous reviewers were constructive and helped to improve this contribution.